\documentclass{aa}

\usepackage{graphicx}
\usepackage{txfonts}
\usepackage{natbib}

\begin{document}

\newcommand{\tmop}[1]{\ensuremath{\operatorname{#1}}}
\newcommand{\pd}[2]{\frac{\partial #1}{\partial #2}}
\newcommand{\HALF}{\frac{1}{2}}
\renewcommand{\vec}[1]{\mathbf{#1}}
\newcommand{\nx}{\hat{\mathbf{e}}_x}
\newcommand{\ny}{\hat{\mathbf{e}}_y}
\newcommand{\nz}{\hat{\mathbf{e}}_z}
\newcommand{\figwidth}{0.75\textwidth}


\title{Aspect ratio dependence in magnetorotational instability
 shearing box simulations} 
 
 \author{G. Bodo\inst{1}
\and         A. Mignone\inst{1,2}         
\and         F. Cattaneo\inst{3} 
\and         P. Rossi\inst{1} 
\and         A. Ferrari\inst{2,3}
}

\offprints{G. Bodo}

 \institute{INAF, Osservatorio Astronomico di Torino, Strada Osservatorio 20,
              Pino Torinese, Italy
 \and
Dipartimento di Fisica Generale, Universit\'a di Torino, 
              Via Pietro Giuria 1, 10125 Torino, Italy
\and 
 Department of Astronomy and Astrophysics, The University of Chicago, 
              5640 S. Ellis ave., Chicago IL 60637, USA}

\date{Received; accepted}

\abstract
{} 
  {We study the changes in the properties of turbulence driven
  by the magnetorotational instability in a shearing box, 
   as the computational domain size 
  in the radial direction is varied relative to the height }
{We perform 3D simulations in the shearing box approximation,
 with a net magnetic flux, 
  and we consider computational domains with different aspect
  ratios}
{We find that in boxes of aspect ratio unity the transport
  of angular momentum is strongly
  intermittent and dominated by channel solutions in agreement with previous work. In contrast, in
  boxes with larger aspect ratio, the channel solutions and the
  associated intermittent behavior disappear.}
{There is strong evidence
  that, as the aspect ratio becomes larger, the characteristics of the
  solution become aspect ratio independent. We conclude that shearing box
  calculations with aspect ratio unity or near unity may introduce spurious effects.}

 
 \authorrunning{Bodo et al.}
 \titlerunning{Aspect ratio dependence in MRI simulations}
\maketitle 
 
\section{Introduction}

Understanding the process of angular momentum transport is one of the
fundamental issues in the physics of accretion disks. Turbulence was 
recognized as the main source of the required enhanced transport \citep{SS73},
 but
its driving mechanism remained unclear until \citet{BH91} proposed the
magnerotational instability (MRI). 
Much of what is presently known about MRI driven turbulence has been
obtained in the shearing box approximation in which only a small periodic
patch of the disk is simulated\citep[see e.g.][]{HGB95, SI01, SS02, B03,
  SITS04, TSKS03, LL07}.  The advantage of this local approach, as opposed to
a global disk simulation, is the possibility of reaching much higher
resolutions at the same computational cost.
The shearing box approach
is of course meaningful only if it  captures the
characteristics of MRI driven turbulence, and of the related angular
momentum transport in a full disk. A critical discussion of the 
validity of the shearing box approximation can be found in a recent work by
 \citet{RU08}  

Ideally one should compare local and global results; however available
global disk simulations \citep[see e.g.][]{H01, HBS01} have relatively low resolution
while simulations with 
adequately high resolution are still beyond 
present capabilities. 
Given the present resources, one should minimally 
check the self consistency of the shearing box
results. The shearing box equations are obtained  as a formal local
expansion in the limit of large radii and small gap \citep{H878}, but there
is no guarantee that the solutions thus obtained satisfy  the
same locality conditions. This should be verified {\it a posteriori}, by
checking, for example, that the properties of the solutions do not depend on
the size of the computational domain.

It useful to distinguish two related issues. One concerns the size of
the computational domain relative to some characteristic length scale
of the problem determined by the physical parameters. For instance, in
the MRI case, the vertical wavelength of the mode of maximum growth
rate depends on the strength of the applied uniform magnetic field.
For a given field strength then, the vertical size of the
computational domain is chosen to contain some multiple of that
wavelength.  In this case, varying the field strength is in some sense
equivalent to varying the (vertical) box size. A related issue
concerns the aspect ratio, i.e the size of the computational domain in
the radial and azimuthal directions relative to the vertical size. The
modes of maximum growth rate do not offer much guidance in this
respect since they only vary in the vertical. Clearly a very thin box
may introduce spurious effects, a very wide box rapidly becomes
computationally expensive. A typical compromise is to adopt boxes with
aspect ratio between the radial and vertical direction of unity.
Presumably, the basis for this choice is that once nonlinear processes
saturate the exponential growth, the typical resulting structures will
have a characteristic size in the radial direction comparable, or in
any case, not much larger than that in the vertical. This being the
case, a computational domain of unit aspect ratio is adequate to
capture the properties of the solution. Strangely, this point has not
received careful consideration and there has been no systematic study
of the dependence of the solutions on the aspect ratio. Furthermore
many studies in computational domains of unit aspect ratio have shown
intermittent behavior associated with the formation and subsequent
disruption of ``channel" solutions \citep{BH92, SI01}. These are the
nonlinear analogues of the exponentially growing linear
modes.  Recently a number of authors have addressed the effect of the
presence or absence of channel solutions in different contexts, for
example, by introducing vertical gravity and stratification
\citep{CK03} or changing the radial boundary conditions \citep{LGJ06,
  URM07, UMR07}.    
Here we note that, interestingly, in the nonlinear regime, the observed channel solution
is not necessarily the one that corresponds to the linear mode of
maximum growth rate. In fact, often, the observed channel solution is
the one whose vertical extent fills the computational domain once. In
this case, it is not a priori obvious what aspect ratio should be
used.  In this paper we address these issues by carrying out a
systematic study of shearing box results as a function of aspect
ratio.  In particular, we want to see if the results observed in boxes
of unit aspect ratio are representative of more extended systems, or
if they display peculiarities induced by an overly constrained
geometry.

\section{Formulation} \label{sec_for} 

We perform a series of 3D, compressible, isothermal numerical
simulations in the shearing box approximation (for a detailed
description of the shearing box model, see \citet{HGB95}). 
The Cartesian coordinates $x$, $y$, $z$ are defined around a fiducial point 
comoving with angular velocity $\Omega$ and correspond, respectively, to the 
radial, azimuthal and vertical directions. 
An equilibrium solution is given by a uniform shear flow, 
$\vec{v} = -q\Omega x\ny$, with $q = 3/2$ for a Keplerian disk,
threaded by a uniform vertical magnetic field, $\vec{B} = B_0\nz$,
in which the density $\rho$ and the pressure $p$ are constant ($\rho = 1$,
$p = c_s^2\rho$).  We have assumed, as it is done by most of the
literature on the subject, the simplest case with no vertical
stratification and gravity. This may limit the applicability to real
disks \citep{RU08}, but it is adequate for our purpose of studying the aspect ratio
dependence. This equilibrium is initially perturbed by 
small amplitude spatially uncorrelated azimuthal velocity perturbations. 
The box has size $L\times 4\times 1$ in the $x$, $y$ and $z$ directions, 
respectively, with $L$ varying from $1$ to $8$.

We take $1/\Omega$ as the unit of time (note however that in the plots
time is in unit of the rotation time $2 \pi / \Omega$),
 and in all the simulations
the sound speed $c_{s}$ and the plasma $\beta=p_{gas}/p_{mag}$ are fixed with values 
of  $4.56$ and $10^{4}$, respectively. With these choices the fastest 
growing MRI mode has a vertical wavelength close to $1/3$.

Computations are carried at both low and high resolutions ($32$ and $128$ zones
per unit length, respectively) on equally spaced grids.
The MHD equations are solved in conservative form using the 
isothermal MHD module available in the PLUTO code \citep{PLUTO,M07}.
The latter is a finite volume, Riemann solver based code in which the
evolution of the magnetic field is carried out using the constrained
transport method of \cite{BS99}.  In the present formulation we do not include explicit viscosity or magnetic diffusivity, and rely solely on numerical dissipation to 
limit the potential unbounded growth of the solutions (for a theoretical discussion of numerical dissipation in Riemann solver based schemes see, for example,  
\cite{T99}; for a discussion of the role of numerical dissipation in turbulent simulations see \cite{GMR07}). 

\begin{figure*}[ht]  
  \centering  
   \includegraphics[width=\hsize]{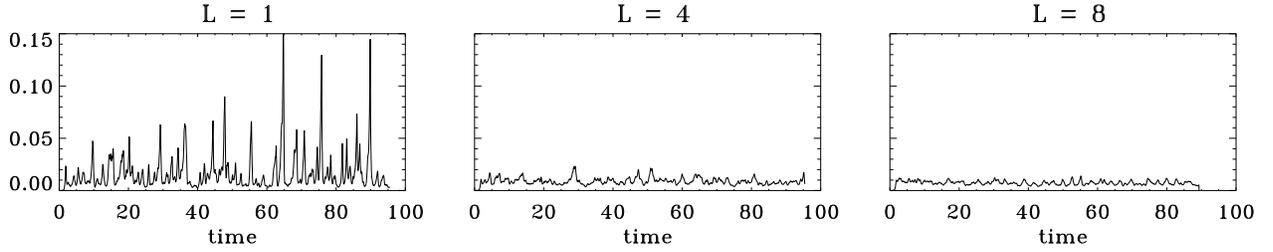}
  \caption{\footnotesize Time histories of the Maxwell stresses averaged over
    the computational box and normalized to the pressure for the low resolution 
    simulations. The three panels correspond to cases with $L=1$, $4$, and $8$. 
    Note that time is in unit of the rotation time.}
  \label{fig:stresses}  
\end{figure*} 
\begin{figure}[ht]
  \centering  
    \includegraphics[width=\hsize]{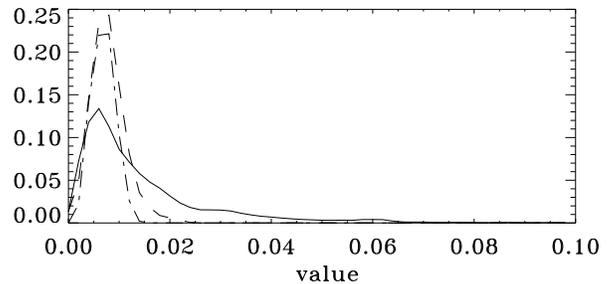}   
  \caption{\footnotesize Probability distribution
    functions of the Maxwell stresses for the same cases as in 
    Figure~\ref{fig:stresses}. The three curves correspond to the three
    values of the aspect ratio:  $L = 1$---solid curve, 
    $L = 4$---dashed curve, and $L = 8$---dot-dashed.}
  \label{fig:pdfstress}  
\end{figure}  
%


\section{Results} \label{sec_lin} 
Our main objective is to study how the properties of the turbulent
solutions vary as the (radial) aspect ratio $L$ is varied from $1$, to
$4$, to $8$. We recall that in simulations with aspect ratio unity, the
angular momentum transport shows an intermittent behavior with
episodes of high transport. These correspond to states, loosely
referred to as ``channel" solutions, in which the velocity and magnetic
perturbations are highly spatially correlated. Strictly speaking, the
channel solutions are exponentially growing exact solutions of the
incompressible shearing box equations in which the velocity and
magnetic field perturbations depend sinusoidally on $z$, have
directions at right angle to each other, and a growth rate determined
by the angle they make to the azimuthal direction \citep{GX94}. More commonly the
term ``channel solutions" is used to describe a behavior with similar
properties to that observed in the exact solutions. It was shown by
\citet{GX94} that the channel solutions, as their amplitudes grow,
become unstable to parasitic instabilities. It is now believed that
the formation of near channel solutions and their subsequent
disruption by the parasitic instabilities are the basis for the
observed intermittent behavior \citep{SI01}. It should be noted that the vertical
wavenumber determines the angle $\gamma_c$ between the direction of
the magnetic perturbations and the azimuthal direction, and hence the
growth rate. 
Although, in principle, channel solutions exist with any
wavenumber, the most commonly observed ones have a vertical extent 
equal to the vertical box size, for which the angle $\gamma_c \approx 23^o$. 
We shall use these results presently.

\begin{figure*}[ht]
  \centering  
  \includegraphics[width=\hsize]{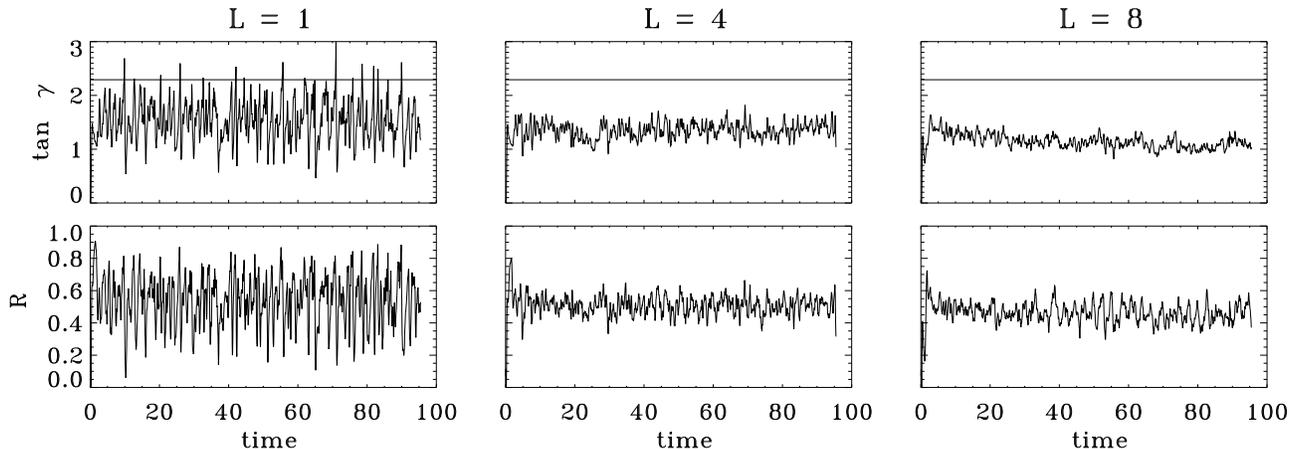}  
  \caption{\footnotesize 
  Time histories of the average slope $\tan \gamma$ (upper row) and correlation coefficient $R$ (lower row) for the same cases as in
  Figure~\ref{fig:stresses}. The straight horizontal lines indicate the theoretical value of $\tan
  \gamma_{c}$ for the exact channel solution with wavelength
  unity. Note that time is in unit of the rotation time.}
  \label{fig:correlation}  
\end{figure*}

\begin{figure}[ht]  
  \centering  
  \includegraphics[width=\hsize]{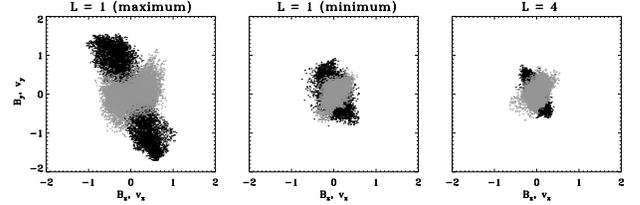}  
  \caption{\footnotesize  Scatter diagrams of horizontal magnetic fluctuations (dark tones) and velocities (lighter tones). The first panel corresponds to the case 
with $L=1$ at an instant near a maximum in the stress. The second panel corresponds to the same case near a minimum. The third panel corresponds to a representative 
instant for the case with $L=4$.}
  \label{fig:scatter}  
\end{figure}

We start our discussion with the low
resolution results and compare three cases with
different aspect ratio. Fig. \ref{fig:stresses}  shows the
the  volume averaged Maxwell stresses 
$\langle w_{xy}\rangle=-\langle B_x B_y\rangle$ as a function of
time. 
In what follows we concentrate on the Maxwell stresses because they give the
largest contribution to the total stresses exceeding the Reynolds
stresses on average by a  factor of about $5$. 
We have normalized the stresses to the pressure, thus the values shown in
the figure correspond  to the $\alpha$ parameter \citep{SS73}. 
The left panel, corresponding to $L = 1$, shows the
characteristic intermittent behavior described above. 
A striking difference appears if we compare this case with the other two 
having larger aspect ratios in which the spikes are absent.
A more quantitative view of the differences between the three cases
is afforded by their respective probability distribution
functions shown in Fig \ref{fig:pdfstress}. 
The case $L = 1$ (solid curve) is quite distinct from the other two with a 
long tail corresponding to the episodes of enhanced transport.
Careful examination  shows that a small tail is still present in the case with 
$L=4$ (dashed curve), and absent in the case with $L=8$ (dot-dashed curve). 
However, we see that the differences between the last two cases are much smaller 
than the differences with the $L = 1$ case, indicating that we have almost reached a
converged behavior. 
Because of the presence of the peaks, the time-averaged value of $\alpha$ for 
$L=1$ exceeds the other two cases by approximately a factor of $2$.
It is natural to assume that the distinctive behavior of the case with $L=1$ 
is due to the presence of channel solutions that are otherwise absent in 
the simulations with larger aspect ratios. 
This can be verified by looking for typical imprints of the channel solutions 
such as, for instance, a high correlation coefficient between directional 
components of the magnetic filed, or a specific relationship between the vertical
wavenumber and the angle between the magnetic field and the azimuthal direction. 
In order to define these quantities it is useful to introduce the idea of a scatter
diagram whereby for each gridpoint the value of $B_y$ is 
plotted as a function of $B_x$, Fig \ref{fig:scatter}. 
For a pure channel solution this procedure would yield a straight line through 
the origin with slope $\tan \gamma_c$. 
In the  general case, the points will not lie on a straight line; still, 
a least-square straight line fit through the points can be used to define 
the average angle, and the spread about this line would give a measure of 
the correlation coefficient $R$. Equivalently, these quantities can be defined by
\begin{equation}
\tan \gamma = -\frac{\langle B_x B_y\rangle}{\langle B_x^2\rangle}, \quad {\rm and} \quad 
R=\frac{\langle B_x B_y\rangle^2}{\langle B_x^2\rangle \langle B_y^2\rangle}.
\label{eqn:coef}
\end{equation}
The time histories of these quantities for different aspect ratios are shown in 
Fig. \ref{fig:correlation}, clearly indicating the presence 
of channel solutions in the case with aspect ratio unity and their absence in 
the other cases.
Comparison between Figures \ref{fig:stresses} and \ref{fig:correlation} 
also confirms that the spikes with high transport correspond, indeed, to 
the formation of channel solutions. 

Differences between the various cases can be further illustrated by
scatter plots of the distribution of $B_{y}$ as a function of $B_{x}$
like those presented in Fig. \ref{fig:scatter}. The two panels on the
left show two examples of distributions for the case $L = 1$, one
corresponding to a maximum of $\langle w_{xy}\rangle$ and the other
corresponding to a minimum. This confirms that in going from the
minimum to the maximum the magnetic field fluctuations increase their
intensity and the $x$ and $y$ components tend to become more
correlated. The velocity fluctuations show a similar behavior, with
almost no correlation in the minimum state. In contrast, for $L = 4$, 
there are no significant variations in the distributions
during the evolution, see Fig. \ref{fig:scatter} where we display only 
one representative example that shows a similarity with the minimum state of 
the case with $L=1$.
The figure also makes apparent why the Reynolds stresses are
significantly lower than the Maxwell stresses. The difference partly 
arises because the intensity of the fluctuations is smaller, 
but to a somewhat larger extent because the velocity fluctuations, 
except during a channel solution phase, remain uncorrelated.
%

\begin{figure}[ht] 
  \centering  
  \includegraphics[width=\hsize]{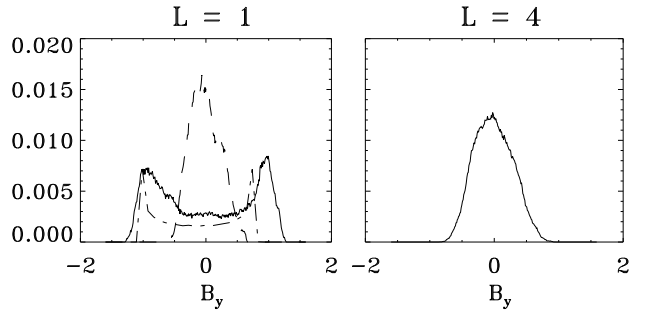}  
  \caption{\footnotesize Probability distribution functions for the
    azimuthal component of magnetic field $B_{y}$. 
	The left panel refers to the case with $L=1$. 
    Solid and dashed lines correspond to the first two panels in 
    Fig. \ref{fig:scatter} while the dot-dashed line corresponds to the 
    exact channel solution. 
    The right panel corresponds to the third panel in Fig. \ref{fig:scatter}.}
  \label{fig:pdfb2}  
\end{figure}

Another property of the distributions not entirely evident from the figure 
is that, while for $L = 4$ and at the minimum of the $L = 1$ case the 
distributions peak at zero value, at the maximum of the $L = 1$ case there are 
two peaks around the largest values of the fluctuations. 
This is clear in Fig. \ref{fig:pdfb2} where we show the probability distribution
function of the azimuthal magnetic field component $B_{y}$. 
The panels show the behavior at an instant near a maximum in the stress 
(solid line) and a minimum (dashed line) for two different aspect ratios.  
In the left panel, we also show (dot-dashed line) the corresponding distribution 
for the channel solution. 
The presence of the double peak distribution in the case with $L = 1$ (left panel) 
can be considered again as an indication of the presence of the channel solution 
around the maxima in the stress, as it can be seen by comparing the solid and the 
dot-dashed lines. 
\begin{figure}[ht]
  \centering  
  \includegraphics[width=\hsize]{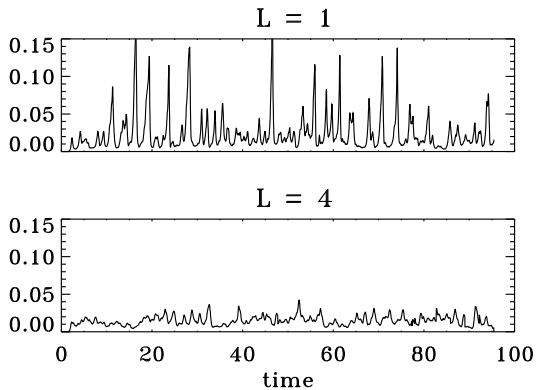}  
  \caption{\footnotesize  
   Time histories of the Maxwell stresses averaged over
   the computational box and normalized to the pressure for the high
   resolution simulations. The two panels correspond to cases with
   $L=1$, and $4$. Note that time is in unit of the rotation time.}
  \label{fig:hr}  
\end{figure}  
Recently, the work of \citet{FR07} has raised some concerns about the dependence of 
numerical studies of MRI turbulence on resolution.
Accordingly, we have repeated the simulations with $L=1, 4$ at a higher resolution 
of $128$ gridpoints per unit length. 
We find the results to be qualitatively the same. 
For example, Fig. \ref{fig:hr} shows the time histories of 
$\langle w_{xy}\rangle$ for the higher resolution simulations. 
The behavior is strikingly similar to that of Fig.~\ref{fig:stresses}, 
with a strong presence of the channel solutions in the $L=1$ case and a 
complete absence of such solutions in the runs with larger aspect ratio. 
We have also checked other indicators, like the probability distribution functions and the scatter diagrams and found 
them to be qualitatively the same as in the cases described above.   
We should note however that quantitatively we do observe a dependence on resolution; in particular the average value of the Maxwell stresses increases with 
increasing resolution (cf. the zero flux case \cite{FR07}). This follows mainly from an increase in the peak values of magnetic fluctuations, which is not surprising 
since the effective magnetic Reynolds number associated with numerical dissipation increases with increasing resolution.

\section{Conclusion} 
Channel solutions represent an highly correlated state for which the
transport is very efficient and are a characteristic feature that strongly
affects the behavior of the angular momentum transport in most of the
shearing box  simulations so far presented in the
literature. There is however the question whether the dominance of
this state could not be due to the constraints imposed by the
typical shearing box approach. Recently, several authors  have
addressed this question considering the effects of vertical
gravity and stratification \citep{CK03}, and of different boundary
conditions \citep{LGJ06, UMR07, URM07}. The general result is that the
channel solution is not robust and disappears whenever one introduces
some modification to the typical shearing box approach and, as a
consequence, the angular momentum transport is less efficient.         

In this paper we have focused our analysis on the effects of a change
of the box aspect ratio and   
the results above show two different behaviors depending on it. 
For aspect ratios close to unity the solutions are strongly
affected by the channel solutions, with frequent episodes of high
correlation and efficient transport. For larger aspect ratios the
channel solutions disappear, the system remains in the state with
lower correlation and the average transport of angular momentum is
correspondingly reduced. Further increase in aspect ratio does not
lead to any significant changes. Thus we conclude that the less
correlated state is more likely to be representative of the extended
system.

It is interesting to speculate why the channel solutions disappear
from boxes with larger aspect ratio. One possibility is to assume that
nontrivial correlations are necessary to form the channel solutions
and that these cannot develop faster than the sound crossing time.
Accordingly, in a large box it takes longer to form the channel
solutions; if the rate at which they are destroyed is fixed in a
sufficiently large box they may never form. We do not believe this is
the correct explanation, since we have found significantly the same
behavior in simulations with a sound speed ten times larger.  Another,
more likely explanation, is that the secondary instabilities
responsible for the destruction of the channel solutions are to some
extent suppressed at small aspect ratios, but can achieve their
maximal growth rate once the aspect ratio is large enough. In support
of this idea, we note that the parasitic instabilities studied by
\citet{GX94} in general have a horizontal size larger than the
vertical size of the channel solution on which they grow. In a box of
unit aspect ratio containing a single channel they may simply be
stable.

Whatever the reason for the differences, clearly, the obvious
conclusion is that studies of turbulence driven by the MRI with net
flux should be conducted in shearing boxes sufficiently large to allow
the solution to develop naturally. The exact domain size depends on
the particular problem and the quantities under consideration.
However, it seem clear that boxes with aspect ratio close to unity
over-emphasize the role of the channel solutions and may lead to
misleading results.

On the other hand our discussion does not put in doubt the fact that
MRI may be adequate to produce the turbulence necessary to support the
required "viscosity" to transport angular momentum.  However, the 
uncertainties discussed by
many authors about the applicability of the shearing box approximation
for the astrophysical problem of accretion disks suggest that the
extension of the shearing box results to the full disk has to be
tested on global simulations with sufficient resolution, that may soon
be feasible.

\section{Acknowledgment}
The authors wish to thank Prof. J. Goodman and the anonymous referee 
for useful comments.  This
work was supported in part by the National Science Foundation
sponsored Center for Magnetic Self Organization at the University of
Chicago.  AF was also partially supported by the US Department of
Energy under grant No. BS23820 to the Center for Astrophysical
Thermonuclear Flashes at the University of 
Chicago.  AM, AF and PR acknowledge support by the Italian Ministro
dell'Universit{\`a} e della Ricerca by PRIN 2005. Calculations were
performed at CINECA (Bologna, Italy) 
thanks to support by INAF.

\end{document}